\documentclass[%
twocolumn,
superscriptaddress,
longbibliography,
nofootinbib,
amsmath,amssymb,
aps,
notitlepage
]{revtex4-1}

\usepackage{graphicx}
\usepackage{dcolumn}
\usepackage{bm}
\usepackage{hyperref}
\usepackage{color}

\usepackage{amsmath}
\usepackage{acronym}
\usepackage{xspace}
\usepackage{subfigure}
\usepackage{longtable} 
\usepackage{multirow}
\usepackage{mathtools}
\usepackage{soul}
\usepackage{orcidlink}
\usepackage{makecell}

\newcommand{\IUCAA}{Inter-University Centre for Astronomy and Astrophysics, Post Bag 4, Ganeshkhind, Pune 411 007, India}
\newcommand{\IPMU}{Kavli Institute for the Physics and Mathematics of the Universe (WPI), The University of Tokyo, 5-1-5, Kashiwanoha, Kashiwa, Chiba 277-8583, Japan}
\newcommand{\ICRR}{KAGRA Observatory, Institute for Cosmic Ray Research, The University of Tokyo, 5-1-5, Kashiwanoha, Kashiwa, Chiba 277-8583, Japan}

\newcommand{\kmsMpc}{\ensuremath{\mbox{km s}^{-1} \,\mbox{Mpc}^{-1}}\xspace}
\newcommand{\Mpcvol}{\ensuremath{\mbox{Mpc}^{-3}}\xspace}
\newcommand{\LCDM}{\ensuremath{\Lambda\mbox{CDM}}\xspace}

\newcommand{\nbar}{\bar{n}}

\newcommand{\dobs}{\delta_{\rm obs}}
\newcommand{\dmiss}{\delta_{\rm miss}}
\newcommand{\nobs}{n_{\rm obs}}

\newcommand{\nbobs}{\bar{n}_{\rm obs}}
\newcommand{\nmiss}{n_{\rm miss}}
\newcommand{\nbmiss}{\bar{n}_{\rm miss}}
\newcommand{\ngrid}{n_{\rm grid}}
\newcommand{\bobs}{b_{\rm obs}}
\newcommand{\bmiss}{b_{\rm miss}}
\newcommand{\brel}{b_{\rm rel}}

\newcommand{\lvk}{\ac{LVK}\xspace}
\newcommand{\gw}{\ac{GW}\xspace}
\newcommand{\emg}{\ac{EM}\xspace}
\newcommand{\bao}{\ac{BAO}\xspace}

\begin{document}

\title{Recovering Large-Scale Clustering of Missing Galaxies for Galaxy--Gravitational-Wave Cross-correlations}

\author{Tathagata Ghosh~\orcidlink{0000-0001-9848-9905}}
\affiliation{\ICRR}

\author{Surhud More~\orcidlink{0000-0002-2986-2371}}
\affiliation{\IUCAA} \affiliation{\IPMU}

\begin{abstract}

We present a framework for completing galaxy catalogs from flux-limited surveys which reconstructs the missing galaxy population while preserving its clustering properties. This is particularly important for cosmological analyses using gravitational-wave (GW) dark sirens without electromagnetic counterparts, such as binary black hole mergers, which constitute the majority of GW detections. As GW detectors probe increasingly larger distances, galaxy catalogs become progressively incomplete owing to survey flux limits. Current state-of-the-art statistical host identification methods typically account for this incompleteness by assuming that the missing galaxies are uniformly distributed in comoving volume. While this approximation can mitigate biases due to incompleteness, it neglects the clustering of galaxies within the cosmic web. Since galaxies and GW sources are both expected to trace the underlying large-scale structure, their spatial distributions are correlated, making a homogeneous reconstruction physically unrealistic, a shortcoming which our framework addresses. As a proof of concept, we apply the method to simulated galaxy catalogs with different flux limits and observational selection functions. Our approach provides a significantly more realistic completion of incomplete galaxy catalogs than the commonly adopted homogeneous-in-comoving-volume approximation, enabling more robust cosmological inference from gravitational-wave dark sirens.

\end{abstract}

\date{\today}

\maketitle

\acrodef{LVK}{LIGO--Virgo--KAGRA}
\acrodef{GW}{gravitational wave}
\acrodef{EM}{electromagnetic}
\acrodef{2PCF}{two-point correlation function}
\acrodef{BAO}{baryon acoustic oscillation}

\section{Introduction}

The use of \gw observations for cosmology has become increasingly promising with the growing number of detected events. By analysing $142$ \gw events from the GWTC-5.0 catalog~\cite{LIGOScientific:2026sit, LIGOScientific:2026wfs}, the \lvk collaboration reported an independent measurement of the Hubble constant,
$H_{0}=71.7^{+9.4}_{-7.5}~\kmsMpc$~\footnote{We quote the median with the $68\%$ symmetric credible interval for the measurement of $H_{0}$ from \gw observations.}~\cite{LIGOScientific:2026uyd}. This result is primarily driven by the bright siren event GW170817, which yields $H_{0}=79.1^{+27.6}_{-12.4}~\kmsMpc$~\cite{LIGOScientific:2025jau} by combining the luminosity distance measured from the GW signal with the redshift inferred from the identification of its host galaxy through \emg observations. In contrast, dark sirens, for which no \emg counterpart is available, require statistical association with galaxies and marginalization over the GW source population~\cite{Mastrogiovanni:2023emh, Gray:2023wgj}. Using this approach, the GWTC-5.0 analysis obtained
$H{0}=68.8^{+14.2}_{-13.2}~\kmsMpc$~\cite{LIGOScientific:2025jau}, demonstrating that dark sirens already provide tighter constraints than those from GW170817. Given their substantially higher detection rate, dark sirens are expected to contribute significantly as probes of the expansion rate with future GW observations.

Current dark siren measurements remain limited by uncertainties in both the GW source population and the galaxy catalogs used for statistical host identification. Although galaxy redshift information substantially improves cosmological inference, present-day GW events typically have large localization volumes, while existing galaxy catalogs are significantly incomplete at cosmological distances. As detector sensitivities improve and next-generation observatories come online, GW localization volumes will shrink, and detections will extend to much higher redshifts. While population modeling will remain an important ingredient of the analysis, the impact of galaxy catalog incompleteness is expected to become an increasingly important systematic.

Galaxy catalogs are inherently flux-limited, causing progressively fainter galaxies to fall below the survey detection threshold at higher redshifts. For example, the GLADE+ catalog~\cite{Dalya:2021ewn}, used in the latest \lvk cosmology analysis, is complete only to approximately $47$ Mpc in the $B$ band. Even after incorporating photometric galaxies, the catalog completeness extends to only $\sim1$ Gpc. Consequently, a substantial fraction of GW events, particularly those at larger distances where most detections are expected, receive little or no direct support from existing galaxy catalogs.

Current LVK analyses account for missing galaxies by supplementing the catalog with a homogeneous prior in comoving volume, thereby correcting the redshift prior for the incompleteness of flux-limited surveys. While this procedure mitigates biases arising from missing galaxies, it suppresses large-scale structure in regions of low completeness, reducing the cosmological information available from high-redshift dark sirens~\cite{Dalang:2023ehp}. In particular, treating unobserved galaxies as uniformly distributed neglects the clustering properties of galaxies that persist beyond the survey detection limit.

In this work, we present a proof-of-concept framework for mitigating galaxy catalog incompleteness by explicitly reconstructing the missing galaxy population while preserving the observed large-scale clustering. We populate regions below the survey detection threshold with synthetic galaxies whose statistical properties are consistent with those of the underlying galaxy distribution. The resulting reconstructed catalog, consisting of both observed and synthetic galaxies, can be incorporated directly into existing LVK dark siren analyses. By explicitly retaining galaxy clustering information, our approach moves beyond homogeneous incompleteness corrections and provides a more informative framework for cosmological inference with dark sirens.

Previous studies have addressed galaxy catalog incompleteness using hierarchical Bayesian reconstructions of the underlying galaxy distribution from flux-limited surveys~\cite{Dalang:2023ehp, Dalang:2024gfk, Leyde:2024tov, Leyde:2025rzk}. These methods jointly model the matter density field, galaxy bias, luminosity function, and survey selection, inferring the complete galaxy catalog through a forward probabilistic model. In contrast, our approach makes no explicit assumptions about the statistical properties of the underlying density field or its correlation function. Instead, we estimate the large-scale overdensity directly from the observed bright galaxy population and populate the unresolved faint galaxies by exploiting the luminosity dependence of galaxy bias. As a result, the reconstructed faint population inherits the realized clustering of the observed galaxy distribution, naturally preserving the large-scale structure encoded in the survey while avoiding the need for an explicit generative model of galaxy clustering. 
While this work was in progress, Ref.~\cite{Barbieri:2026fdy} has independently also proposed populating incomplete catalogs by distributing missing galaxies according to an assumed model for the two-point correlation function, reporting substantial improvements in dark siren $H_{0}$ measurements. While their method assumes a given form of correlation function, we utilize an astrophysically motivated bias prescription in order to find the missing galaxies.

The paper is structured as follows. Sec.~\ref{sec:method} describes the methodology for reconstructing the unobserved regions of galaxy catalogs below the survey magnitude limit. In Sec.~\ref{sec:gal_cat}, we describe the construction of a volume-limited mock galaxy catalog with different selection criteria used to define the observed galaxy samples. We then outline the reconstruction procedures used to populate the unobserved regions corresponding to the selection criteria for selecting the observed galaxies in Sec.~\ref{sec:gal_miss}. In Sec.~\ref{sec:results}, we compare the clustering properties of the true and reconstructed missing galaxy populations. Finally, in Sec.~\ref{sec:conclusion}, we summarize our findings and discuss future directions for applying this framework to real data.

\section{Methodology} \label{sec:method}

In this section, we develop a framework for reconstructing the spatial distribution of galaxies below a survey detection threshold using the observed galaxy population. Our objective is to populate the missing galaxy population while preserving its large-scale clustering, thereby producing a statistically complete catalog suitable for cosmological analyses. The large-scale clustering of galaxies is commonly characterized by the two-point correlation function, $\xi(r)$, which measures the excess probability relative to a random Poisson distribution of finding a pair of galaxies separated by a comoving distance $r$,
\begin{align}
    \delta P = \nbar[1+\xi (r)] dV \,,
\end{align}
where $\bar{n}$ is the mean galaxy number density and $dV$ is the infinitesimal volume element surrounding the second galaxy. Equivalently, the correlation function is the two-point autocorrelation of the overdensity field,
\begin{align}
    \xi(r) \equiv \langle \delta(\mathbf{x})\delta(\mathbf{x}+\mathbf{r})\rangle\,,
\end{align}
where the overdensity is defined as
\begin{align}
    \delta(\mathbf{x})=\frac{\rho(\mathbf{x})}{\bar{\rho}}-1,
\end{align}
with $\rho(\mathbf{x})$ and $\bar{\rho}$ denoting the local and mean matter densities, respectively.

Galaxies are biased tracers of the underlying matter distribution. In the linear bias approximation, the galaxy overdensity is related to the matter overdensity through
\begin{align}
    \delta_g=b_g\,\delta_m,
\end{align}
where $b_g$ is the effective galaxy bias, which may depend on scale and redshift~\cite{Coles:2007be}. Consequently, galaxy populations with different biases trace the same underlying large-scale structure while exhibiting different clustering amplitudes.

To develop and validate our reconstruction framework, we begin with a simulated galaxy catalog covering the full sky. Observational selection criteria, described in Sec.~\ref{sec:gal_cat}, are then applied to divide the galaxies into observed and missing populations. Their overdensity fields in a given redshift slice are defined as
\begin{align} \label{eq:delta_obs}
    \dobs(z) = \frac{\nobs(z)}{\bar{n}_{\rm obs}(z)} - 1\,,
\end{align}
and
\begin{align} \label{eq:delta_miss}
    \dmiss(z) = \frac{\nmiss(z)}{\nbmiss(z)} - 1\,,
\end{align}
respectively, where $\nobs$ and $\nmiss$ denote the number densities of observed and missing galaxies, respectively. The corresponding mean number densities are given by $\nbobs$ and $\nbmiss$.
The central assumption of our reconstruction is that the observed and missing galaxy populations trace the same realization of the underlying matter density field, differing only in their effective linear bias. Under this assumption, the overdensity of the missing galaxies can be related to that of the observed galaxies through
\begin{align}
    \dmiss(z)
    =
    \brel(z)\,
    \dobs(z),
    \label{eq:rel_delobs_delmiss}
\end{align}
where
\begin{align}
    \brel(z)=\frac{\bmiss(z)}{\bobs(z)}
\end{align}
is the relative bias between the missing and observed populations.
Here, $\bobs$ and $\bmiss$ are biases of observed and missing galaxies, respectively.

Galaxies inherit their large-scale clustering bias from the dark matter halos that host them. Since the masses of individual halos are generally not directly observable, galaxy bias must instead be inferred from observable galaxy properties. For example, at low redshift, the relative clustering amplitudes of galaxies in different luminosity bins can be used to determine their relative biases. These empirical relations can then be extrapolated to higher redshifts, where the fainter galaxy population is not directly observable.

However, in this paper, we carry out an averaging of the halo bias of our galaxy population in order to determine $\bmiss$ and $\bobs$. Thus,
\begin{align}
    b(z) = \frac{1}{N}\sum_{i=1}^N b(M_i, z)
\end{align}
We adopt the halo bias mass relation of \cite{Tinker_2010} in order to compute this bias. Thus, we assume that the halo bias and the number density of galaxies to be populated will be known a priori for the purpose of this paper.

Substituting Eqs.~\eqref{eq:delta_obs} and~\eqref{eq:delta_miss} in
Eq.~\eqref{eq:rel_delobs_delmiss}, the number density of the missed galaxies can
be written as
\begin{align} \label{eq:n_miss}
    \nmiss = \nbmiss \left(1 + \brel \dobs\right) \,.
\end{align}
The knowledge of $\nbmiss$, $b_{\rm rel}$, and $\dobs$ thus allows us to estimate the number density field of the missed galaxies, thereby reconstructing the underlying distribution of the fainter galaxy population. Note that we use the overdensity of the observed bright galaxy population directly in order to reconstruct the missing galaxy population. As this is a proof-of-concept study, we compute the relative bias of the missing galaxies directly from the simulation, instead of utilizing the lower redshift galaxies. 

\section{Mock Galaxy Catalog} \label{sec:gal_cat}

We demonstrate the framework described in Sec.~\ref{sec:method} using a simulated galaxy catalog constructed from dark matter halos in the Big MultiDark Planck (BigMDPL) cosmological N-body simulation~\citep{Klypin:2014kpa}, available through the CosmoSim database.\footnote{The database is publicly available at \url{https://www.cosmosim.org/}. The same simulated catalog has also been used in our earlier studies~\cite{Ghosh:2023ksl, Ghosh:2025qwc}.} The BigMDPL simulation follows the evolution of the matter density field in a flat \LCDM cosmology with Hubble parameter $h = H_{0}/(100\ \kmsMpc) = 0.6777$, matter density parameter $\Omega_{m}=0.307$, amplitude of matter fluctuations $\sigma_{8}=0.823$, and scalar spectral index $n_{s}=0.96$. The simulation evolves $3840^{3}$ collisionless particles in a periodic cubic volume of comoving side length $2.5\,h^{-1}\,{\rm Gpc}$, corresponding to a particle mass resolution of $2.359\times10^{10}\,h^{-1}M_{\odot}$.

We first construct a complete mock galaxy catalog by selecting all well-resolved dark matter halos with masses $M_{h} \geq 10^{12}\,h^{-1}M_{\odot}$.\footnote{Here, ``complete'' refers to the absence of any flux limit, although the catalog remains volume-limited.} For simplicity, we assume that every halo hosts only a single central galaxy located at its center. This results in a mock catalog containing $\sim 6.68\times10^{7}$ galaxies, which we refer to as \textit{Catalog A}. To emulate an observational selection, we classify only galaxies hosted by halos with masses $M_{h} \geq 10^{13}\,h^{-1}M_{\odot}$ as observed, yielding an observed sample of $\sim7.35\times10^{6}$ galaxies. This catalog is intentionally idealized, as the entire simulation box is treated as the survey volume with periodic boundary conditions, and the halo masses used to define the observed sample are assumed to be known, although they are not directly accessible in real observations.

To construct a catalog with a more realistic survey geometry, we further restrict Catalog A to galaxies lying within a sphere of diameter equal to the simulation box size and centered on the midpoint of the box. The resulting catalog, hereafter referred to as \textit{Catalog B}, contains $\sim3.85\times10^{6}$ galaxies. We place an observer at the center of the sphere and compute the sky coordinates of each galaxy. Galaxy redshifts are then assigned from their comoving distances, assuming the same cosmological parameters used in the BigMDPL simulation. For simplicity, we neglect both redshift uncertainties and redshift-space distortion effects.

Finally, we construct a more realistic flux-limited galaxy sample by assigning apparent magnitudes to the galaxies in the complete mock catalog. The absolute magnitude of each galaxy is estimated using the calibrated relation between galaxy luminosity and the maximum circular velocity of its host dark matter halo from Ref.~\cite{2011ApJ...742...16T}. To account for the intrinsic scatter in this relation, we add Gaussian random noise with zero mean and a standard deviation of $0.5$~\cite{2011ApJ...742...16T} to the absolute magnitudes. The apparent magnitudes are then computed from these scattered absolute magnitudes using the corresponding luminosity distances, assuming the same cosmology adopted in the BigMDPL simulation.

We apply a fiducial apparent magnitude threshold of $m_{r}\leq 17.77$, typical of the Sloan Digital Sky Survey~\cite{SDSS:2002jsq}, which yields a flux-limited mock galaxy catalog, which we hereafter refer to as \textit{Catalog C}. This introduces a more realistic observational selection than the simplified criteria adopted for Catalog A (and Catalog B). 
This catalog contains the same total number of galaxies as Catalog B; however, the number of observed galaxies differs, with $\sim 7.23 \times 10^{5}$ galaxies identified as observed.

In summary, we employ three realizations of the galaxy catalog in our analysis in increasing order of realism, which we refer to throughout the paper as Catalogs A, B, and C.

\begin{itemize}
    \item \textbf{Catalog A} is a mass-selected catalog constructed using all dark matter halos in the full simulation box that satisfy the chosen mass threshold.

    \item \textbf{Catalog B}  applies the same mass selection as Catalog A, but is restricted to the survey-like spherical region.

    \item \textbf{Catalog C} is a flux-limited catalog obtained by applying an apparent-magnitude threshold to all galaxies within the spherical region (without imposing a mass cut, unlike Catalogs A and B), thereby introducing incompleteness that mimics observational surveys.

\end{itemize}

\begin{figure*}
    \centering
    \includegraphics[scale=0.47]{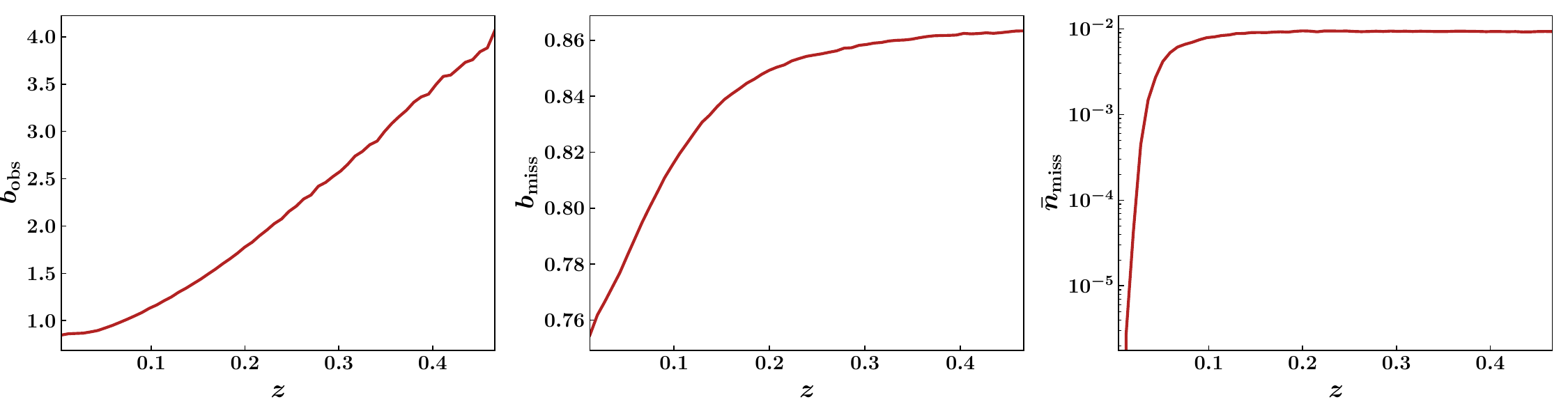}
    \caption{Redshift dependence of the observed galaxy bias, $\bobs$, missing galaxy bias, $\bmiss$, and the mean number density of missing galaxies, $\nbmiss$, for Catalog C, computed using $60$ bins of uniform redshift width.}
    \label{fig:bias_nbar_catC}
\end{figure*}

\section{Reconstructing Missing Galaxies} \label{sec:gal_miss}

In this section, we demonstrate the fundamental premise of the proposed methodology. To focus on the conceptual framework, we assume that the halo bias and the number density of galaxies required to populate the missing galaxy population are known. We compute them directly from the halos in the underlying simulation as discussed at the end of Sec.~\ref{sec:method}. In real galaxy catalogs, the corresponding halo information is not directly accessible, but can be obtained by analyzing galaxies at lower redshifts. These limitations and possible prescriptions for application to real data are discussed in Sec.~\ref{sec:conclusion}.

\subsection{Catalog A} \label{sec:misscatA}

We consider the observed galaxy sample based on central galaxies in dark matter halos with masses $\geq 10^{13}~M_{\odot}$. Since we do not use a light-cone simulation and our selection is defined by a fixed halo mass threshold, the biases and the number density associated with the observed and missing galaxy populations are constant. These quantities do not change with the assigned observed redshift based on the line-of-sight comoving distance. To determine the number density, we use the halo masses of the observed and missing populations. For computing the bias, we average the halo bias over these populations. This yields $\bobs \simeq 1.442$ for the observed galaxies and $b_{\rm miss} \simeq 0.902$ for the missing galaxies. The number density $\nbmiss$ of missing galaxies with masses $\leq 10^{13}~M_{\odot}$, is $\nbmiss\simeq.004~\Mpcvol$. 

Once these quantities are determined, we calculate the overdensity field $\dobs$ from the observed set of galaxies. To do so, we divide the simulation box into a three-dimensional Cartesian grid with $\ngrid$ cells along each dimension, resulting in a total of $\ngrid^{3}$ grid cells. We consider two choices, $\ngrid=256$ and $512$, with a resolution of $9.8h^{-1}{\rm Mpc}$ and $4.9h^{-1}{\rm Mpc}$, respectively. The observed overdensity $\dobs$ is then calculated for each grid cell by utilizing the number of observed galaxies in that cell and the expected number obtained from the mean number density of
observed galaxies in the simulation box.

After all the quantities appearing on the right-hand side of Eq.~\eqref{eq:n_miss} are computed, we evaluate the corresponding expression to estimate the number of missing galaxies. We draw a Poisson deviate from this average expectation, and within each grid cell, we distribute these galaxies randomly to reconstruct the missing component of the galaxy catalog. We compare the statistical properties of the truly missing galaxies with the reconstructed population by computing their respective autocorrelation functions in Sec.~\ref{sec:results}.

\subsection{Catalog B}

The procedure used to reconstruct the missing galaxy population for Catalog B is largely the same as that adopted for Catalog A, with a minor modification arising from the spherical geometry of the galaxy catalog. Since the selection criteria for observed galaxies are identical to those in Catalog A, the quantities $\bobs$, $\bmiss$, and $\nbmiss$ remain unchanged. As described in Sec.~\ref{sec:misscatA}, we randomly distribute the missing galaxies within each grid cell to construct the missing component of the galaxy catalog. Note that when calculating the number of missing galaxies in each grid cell and populating them, we account for grid cells that are partially covered, particularly at the boundary, by the spherical survey volume that lies inside or outside of it. Otherwise, we employ the same three-dimensional Cartesian grid for the entire simulation box. A similar autocorrelation function test as that considered for Catalog A is also carried out for Catalog B.

\begin{figure}
    \centering
    \includegraphics[scale=0.53]{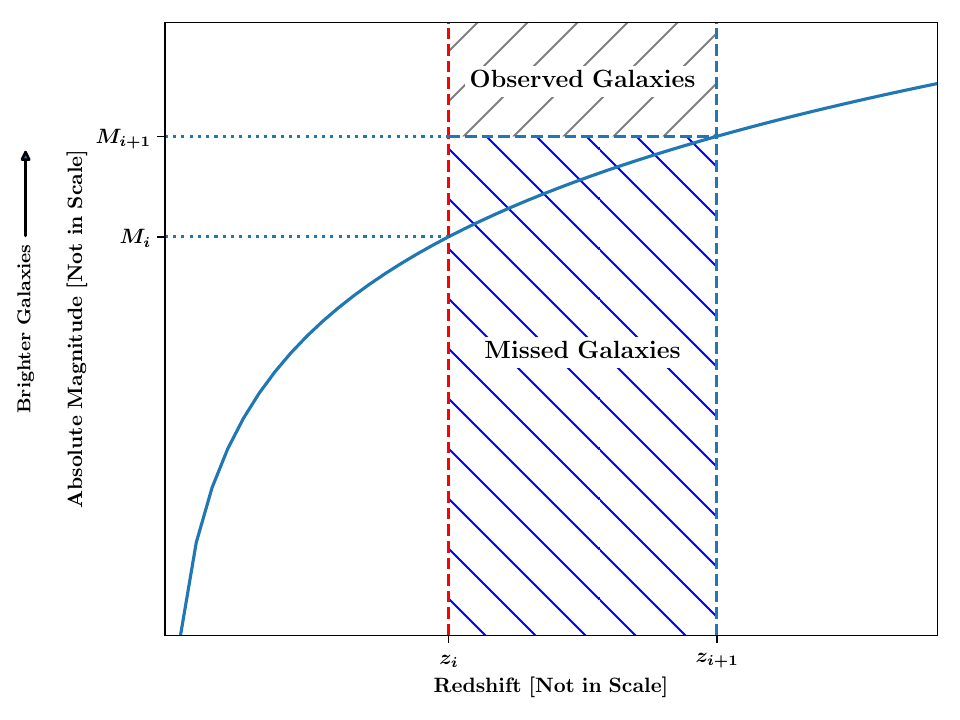}
    \caption{Schematic illustration of the galaxy selection in the $i$th redshift bin. Galaxies with $M < M(z_{i+1})$ are considered observed and are shown with a lightly hatched shaded region, while those with  $M \geq M(z_{i+1})$ are treated as missing and are shown with a densely hatched shaded region. Galaxies with $M(z_{i+1})\leq M <M(z_{i})$ are excluded to ensure a consistent implementation of the selection.}
    \label{fig:illustration}
\end{figure}

\begin{figure*}
    \centering
    \includegraphics[scale=0.58]{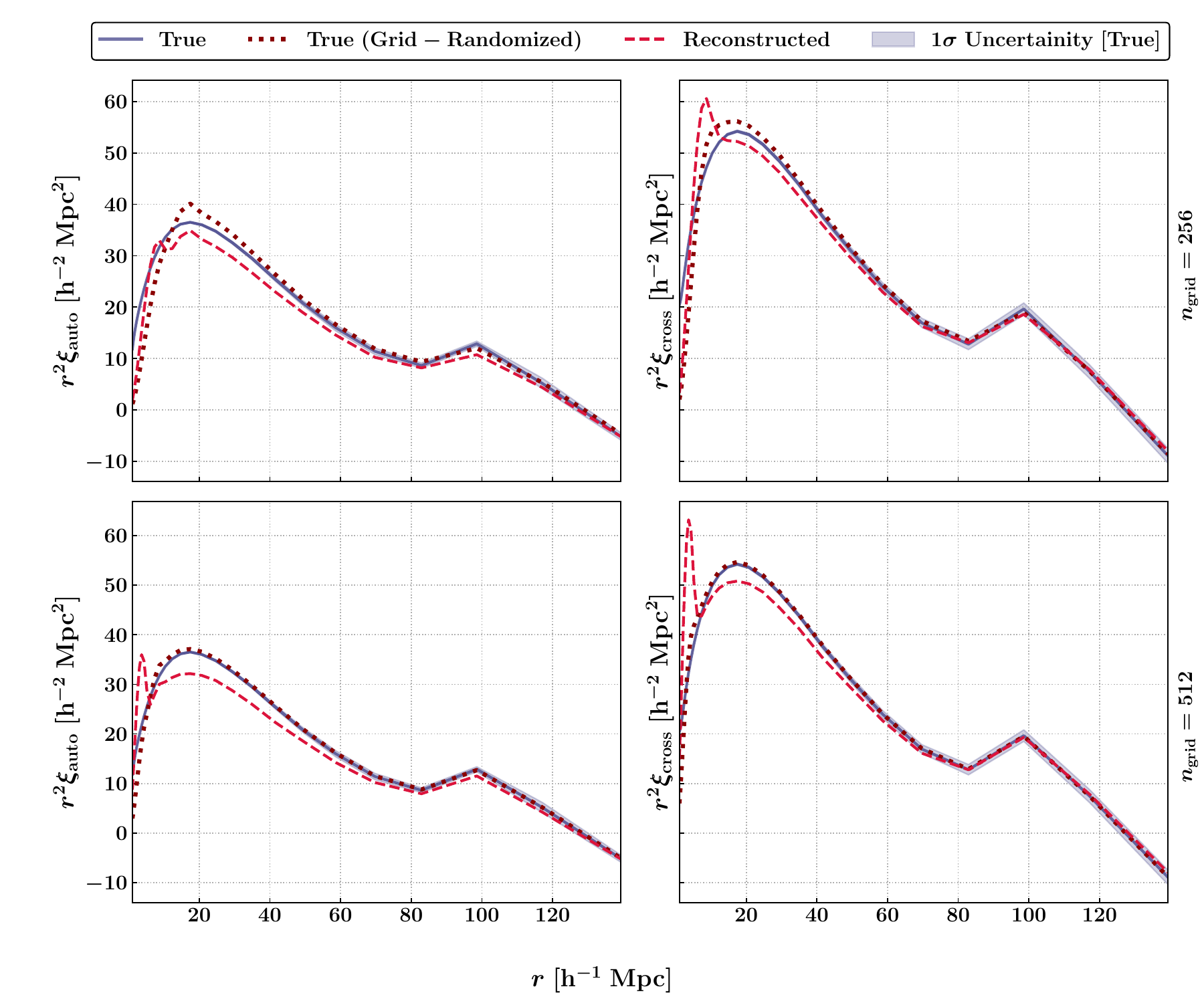}
    \caption{Comparison of the auto- and cross-correlation functions of the truly missing galaxies, their randomization within the corresponding voxels, and the reconstructed missing galaxies for Catalog A. The top row shows the results for $\ngrid=256$, with the autocorrelation in the left panel and the cross-correlation with the observed population in the right panel. The bottom row shows the corresponding results for $\ngrid=512$.}
    \label{fig:xi_catA}
\end{figure*}

\begin{figure*}
    \centering
    \includegraphics[scale=0.58]{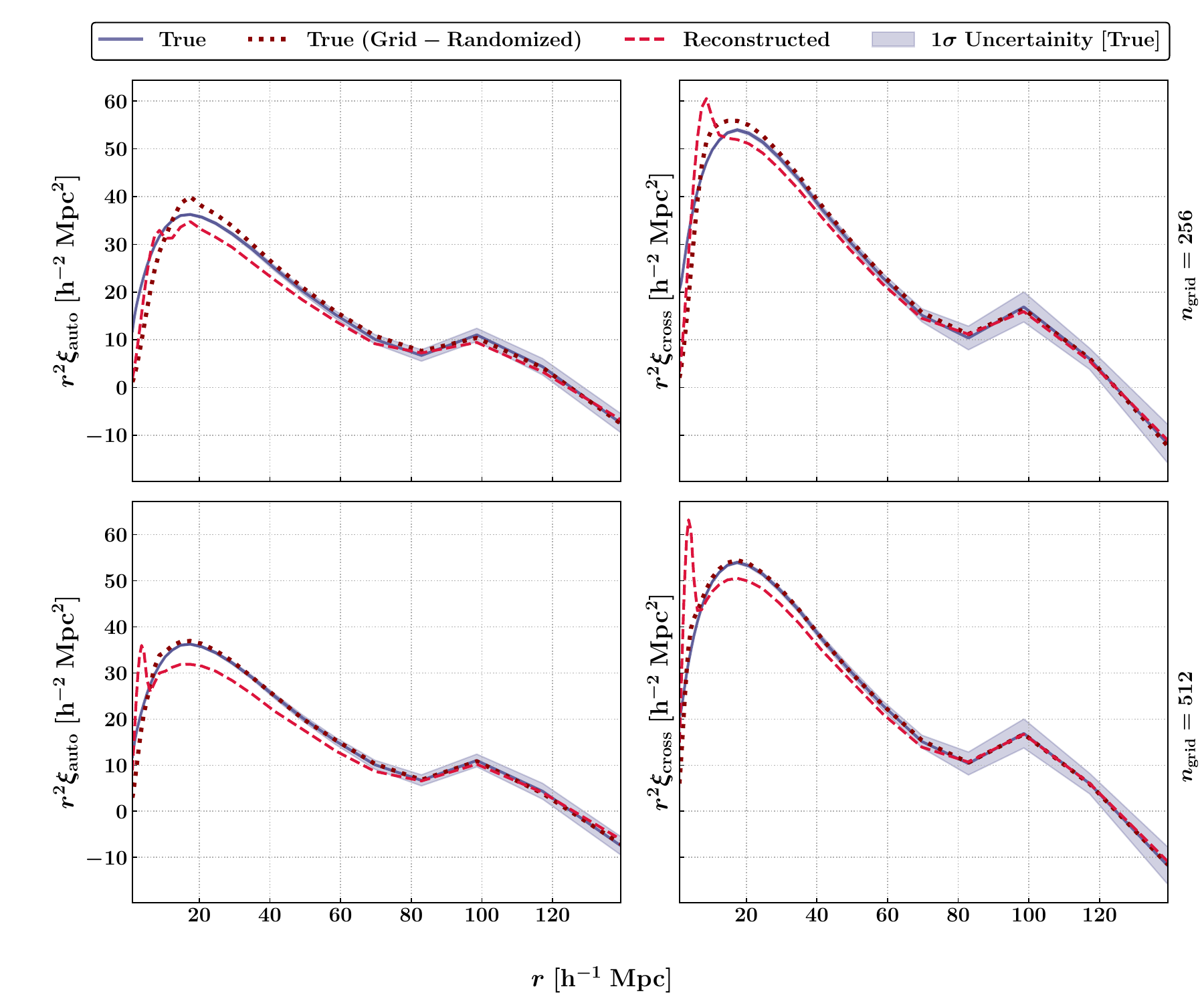}
    \caption{Same as Fig.~\ref{fig:xi_catA}, but using Catalog B.}
    \label{fig:xi_catB}
\end{figure*}

\begin{figure*}
    \centering
    \includegraphics[scale=0.58]{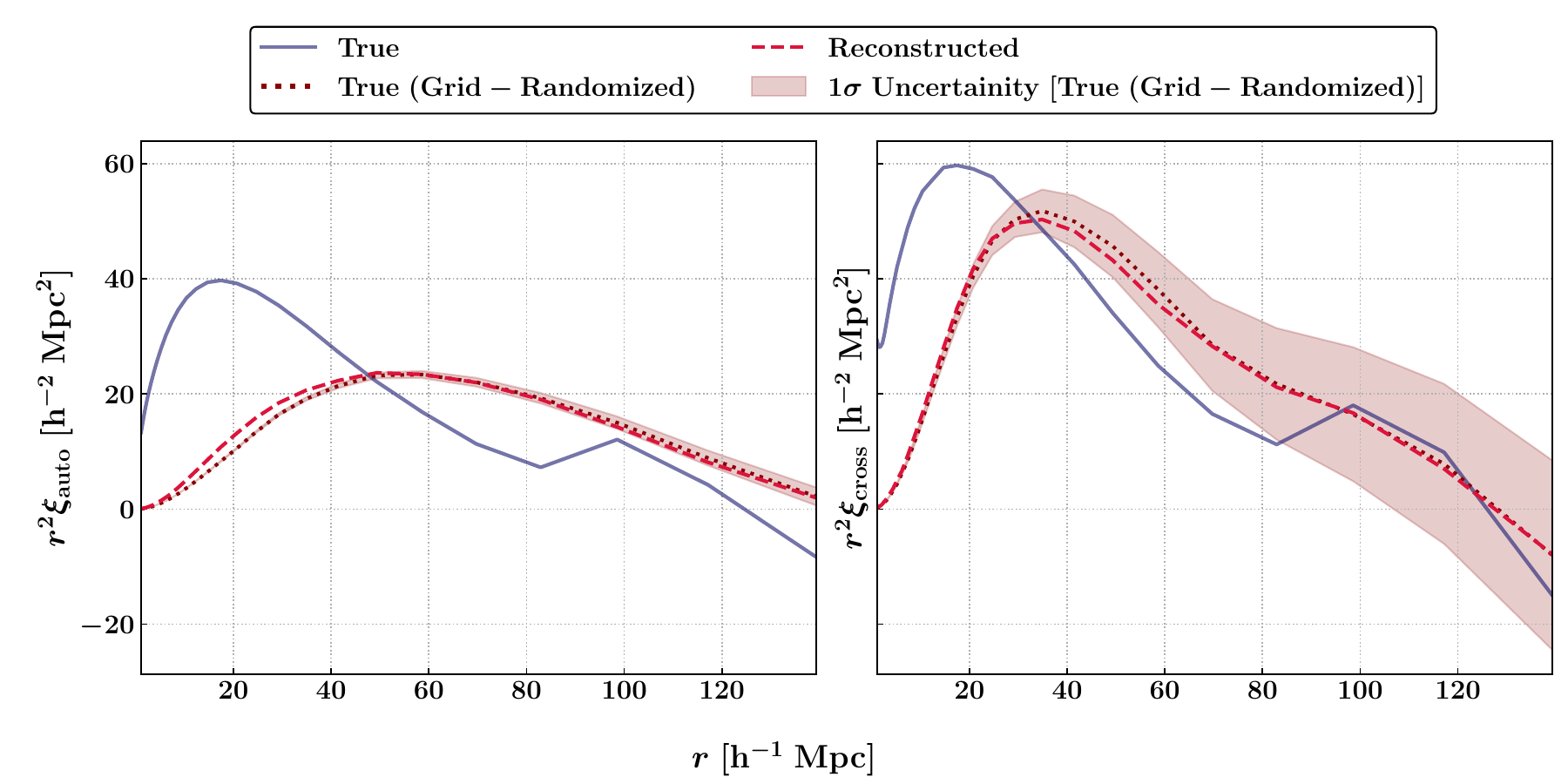}
    \caption{Same as Fig.~\ref{fig:xi_catA} (or Fig.~\ref{fig:xi_catB}), but for Catalog C with volume elements defined using $50$ uniform redshift bins and HEALPix pixelization with $n_{\rm side}=10$ instead of a Cartesian grid.}
    \label{fig:xi_catC}
\end{figure*}

\subsection{Catalog C} \label{sec:gal_miss_C}

Catalog C features an increasing realism of a selection based on an apparent
magnitude cut, $m_{r}\leq 17.77$ rather than a halo mass threshold. Therefore,
we have to consider a slightly different procedure compared to Catalogs A and B.
The selection cut implies that $\bobs$, $\bmiss$, and $\nbmiss$ become redshift
dependent, as shown in Fig.~\ref{fig:bias_nbar_catC}. Therefore, we cannot adopt
the same Cartesian grid–based voxel-wise treatment as in the previous cases, as
it does not account for the redshift dependence of these quantities.

For this catalog, we adopt uniform binning in redshift and HEALPix (Hierarchical
Equal Area isoLatitude Pixelization)-based pixelization~\cite{Gorski:2004by,
Zonca:2019vzt} of the sky to divide it into equal-area pixels at a chosen resolution, defining three-dimensional volume elements. E
ven within the $i$th redshift bin $[z_{i}, z_{i+1}]$, the apparent magnitude limit translates into an absolute magnitude threshold which is not constant. Therefore, conservatively we consider galaxies with absolute magnitude $M < M(z_{i+1})$ at the higher end of
the redshift bin. A schematic illustration of this selection is shown in Fig.~\ref{fig:illustration}. This effectively results in a selection analogous to a fixed halo mass threshold within each redshift bin.

We use $\bobs$, $\bmiss$, and $\nbmiss$ evaluated at $z_{i+1}$ to populate the missing population corresponding to $M \geq M(z_{i+1})$. To implement this consistently, we exclude galaxies with absolute magnitudes in the range $M(z_{i+1})\leq M <M(z_{i})$, even though they are present in the catalog. Once the average number of missing galaxies is estimated for each volume element, we draw a Poisson deviate from this average expectation and distribute the synthetic galaxies uniformly within the cell all over the sky. Finally, we also assess the performance of the reconstruction for Catalog C by comparing the autocorrelation functions, as presented in Sec.~\ref{sec:results}.

\section{Statistical Assessment of the Reconstructed Galaxy Population} \label{sec:results}

We reconstruct the faint missing galaxy populations for all three catalogs (see Sec.~\ref{sec:gal_cat}) following the procedure described in Sec.~\ref{sec:gal_miss}. 
To validate our proposed methodology, we compare the autocorrelation of the reconstructed missing galaxies with that of the true missing galaxies.
We further assess the clustering of the reconstructed missing galaxies relative to the observed galaxy population by comparing their cross-correlation with the observed population to that of the true missing galaxies.
We expect the reconstructed population of galaxies to follow the clustering of the missing galaxies on scales larger than the resolution of the cell. Since we randomly distribute the missing galaxies within a cell, it will miss structure on smaller scales. To enable an appropriate comparison due to this lack of structure, we also generate a cell-randomized realization of the true missing galaxies, in which galaxy positions are randomized within each grid cell, which effectively washes out small-scale clustering while preserving the large-scale clustering. The autocorrelation of these cell-randomized true galaxies is also compared with that of the reconstructed missing galaxies, along with their cross-correlation with the observed galaxy population. To compute the autocorrelation and cross-correlation functions considered in each case, we adopt the Landy--Szalay estimator~\cite{1993ApJ...412...64L}, defined as
\begin{align}
    \hat{\xi}_{\rm LS}
= \frac{D_1D_2-D_1R_2-D_2R_1+R_1R_2}{R_1R_2} \,,
\end{align}
where $D_1D_2$ denotes the number of pairs between the two data catalogs, $D_1R_2$ and $D_2R_1$ denote the corresponding data--random pairs, and $R_1R_2$ denotes the number of pairs between the two random catalogs. For the autocorrelation, the two data catalogs are the same, i.e., $D_1 = D_2$, and likewise for the random catalogs, $R_1 = R_2$. For the cross-correlation, the
two data catalogs correspond to the two distinct galaxy populations being correlated.

In the left panel of Fig.~\ref{fig:xi_catA}, we compare the autocorrelation functions of the truly missing galaxies with those of the reconstructed missing galaxies for two grid configurations, with $256$ and $512$ cells along each dimension (i.e., $256^{3}$ and $512^{3}$ total cells). Since the reconstructed missing galaxies are distributed randomly within their corresponding voxels, we also compare their autocorrelation with that of the truly missing galaxies after randomization within the same voxels, which removes small-scale clustering. 
The uncertainty regions shown in the left panel of Fig.~\ref{fig:xi_catA} are estimated using a jackknife resampling procedure. The catalog (equivalent to the simulation box) is divided into $10^{3}$ equal sub-volumes of side length $250\,{\rm Mpc}$, and the autocorrelation function is recomputed after excluding one sub-volume at a time. The variance is estimated from the resulting ensemble of jackknife realizations using the standard unbiased jackknife estimator~\cite{10.1214/aos/1176345462}. 
As expected, on small scales the reconstructed population loses the sub-voxel clustering of the truly missing galaxies, but is qualitatively similar to that of the true missing population after being randomized within the voxel resolution.
On larger scales, the reconstructed missing galaxies exhibit clustering amplitudes close to those of the true missing population, with a median fractional residual of $\sim 9\%$ for $r \gtrsim 10$~Mpc. In particular, the clustering signal at scales $\gtrsim 80$~Mpc, including the \bao scale, is reasonably well recovered in both grid configurations.

The right panel of Fig.~\ref{fig:xi_catA} shows the corresponding cross-correlation of the observed galaxies with the truly missing, reconstructed missing, and voxel-randomized true missing galaxies. The uncertainty regions shown here are estimated following the same jackknife procedure described above. The reconstructed population recovers the cross-correlation of the truly missing galaxies to within a median fractional residual of $\sim 5\%$ for $r \gtrsim 10$~Mpc, and agrees with the true population within the estimated jackknife uncertainties at large scales. 
We note that the jackknife uncertainties on the autocorrelation are noticeably narrower than those on the cross-correlation. This is because the number of missing galaxies greatly exceeds that of the observed galaxies. The autocorrelation of the missing population therefore benefits from a much larger number of galaxy pairs than its cross-correlation with the smaller observed population, resulting in smaller uncertainties.

For Catalog B, the selection criterion for observed galaxies is identical to that in Catalog A. Consequently, the reconstructed missing galaxies are expected to exhibit similar statistical properties despite the different geometries, as shown in Fig.~\ref{fig:xi_catB}.
We consider two choices of $\ngrid$ to demonstrate that the selection criteria, rather than the underlying geometry, primarily determine the statistical properties, and to assess the impact of the reconstruction grid resolution on the preservation of small-scale clustering.
The jackknife uncertainties on the correlation functions of the true population are slightly broader for Catalog B than for Catalog A, owing to its smaller survey volume. Since the spherical volume of Catalog B is smaller than the full simulation box by a factor of $\pi/6$, the corresponding increase in the jackknife uncertainty is expected to be approximately the inverse of this factor, consistent with our results.

For Catalog C, we show the corresponding comparison in Fig.~\ref{fig:xi_catC}. Note that the implementation of the reconstruction differs from that of the other two catalogs, as discussed in Sec.~\ref{sec:gal_miss_C}. In this study, we adopt $n_{\rm side}=10$ for pixelating the sky and use $50$ bins of equal width in redshift.
We also assess the consistency of the reconstruction by estimating the uncertainty regions shown in Fig.~\ref{fig:xi_catC} using the same jackknife resampling procedure as for Catalogs A and B. In this case, however, the jackknife realizations are generated from the truly missing galaxies after randomization within their corresponding volume elements, rather than from the original truly missing galaxy population, thereby providing a consistent reference for comparison with the reconstructed galaxies.
The autocorrelation of the reconstructed galaxies agrees well with that of the randomized missing galaxies within the corresponding volume elements, with the median fractional residual improving from $\sim10\%$ for $r\gtrsim10$~Mpc to $\sim3\%$ for $r\gtrsim40$~Mpc. However, it does not entirely recover the clustering of the truly missing galaxies.
This behavior is entirely driven by the resolution of our HEALPix grid. Given that $\sim 98\%$ of galaxies are missing in this catalog, a higher-resolution grid can result in low, and in some cases zero, galaxy counts per pixel–redshift bin. Consequently, placing the majority of missing galaxies randomly within each pixel–redshift bin significantly suppresses the clustering signal. This is nevertheless an improvement over the assumption of uniformly distributed galaxies in a comoving volume that is assumed in the standard analyses. This currently represents the optimal level of agreement achievable within the reconstruction framework. The clustering of the reconstructed galaxies closely matches that of the truly missing galaxies randomized within each pixel–redshift bin, which defines the expected limit of the reconstruction.

Similarly, the cross-correlation of the reconstructed population with the observed galaxies shows good agreement with that of the voxel-randomized true population, with a median fractional residual of $\sim7\%$ over all scales, improving to $\sim3\%$ for $r\gtrsim10$~Mpc. As with the autocorrelation, the reconstructed cross-correlation does not fully recover that of the truly missing galaxies. However, the disagreement is smaller, with a median fractional residual of $\sim36\%$ for $r\gtrsim10$~Mpc, compared to $\sim70\%$ for the autocorrelation. This is likely because the observed galaxy population is the same in both cross-correlations, thereby reducing the impact of suppressed clustering of the missing population. Similar to Catalogs A and B, the jackknife uncertainty on the cross-correlation is larger than that on the autocorrelation for Catalog C.

\section{Conclusion} \label{sec:conclusion}

In this study, we have demonstrated a proof-of-concept framework for reconstructing the missing galaxy population in flux-limited catalogs while preserving their large-scale clustering. This offers a more realistic alternative to the homogeneous assumption adopted in current \lvk dark siren analyses.
Our main results can be summarized as follows:

\begin{itemize}
    \item [$-$] We demonstrated a method to reconstruct the galaxy population below a survey’s detection threshold by relating the overdensity of missing galaxies to that of observed galaxies through their relative bias. This approach assumes that both populations trace the same underlying matter density field.

    \item [$-$] We validate the framework using simulated galaxy catalogs of increasing observational realism: a halo mass-selected catalog spanning the full simulation box (Catalog A), the same selection restricted to a survey-like spherical geometry (Catalog B), and a flux-limited catalog constructed using an SDSS-like apparent magnitude threshold (Catalog C).

    \item [$-$] We assess the fidelity of the reconstruction in two ways. First, we compare the autocorrelation function of the reconstructed missing galaxies with the autocorrelation functions of the true missing population and a randomized version of the true missing population, with galaxy positions randomized within the adopted voxel resolution. Second, we compare the cross-correlation of the observed galaxy population with the true missing population, the reconstructed missing galaxies, and the randomized true missing population. Both comparisons are performed using the Landy--Szalay estimator with uncertainties estimated from jackknife resampling.

    \item [$-$] For Catalogs A and B, the reconstructed missing galaxies recover the clustering of the true missing population to within $10\%$ on small scales, and to within the estimated uncertainties on large scales, including the \bao feature.

    \item [$-$] For Catalog C, the recovered clustering is limited by the resolution of the HEALPix--redshift binning given the high level of incompleteness ($\sim98\%$ missing galaxies), but still represents a marked improvement over the assumption of a homogeneous missing population.

\end{itemize}

Although we have adopted several simplifying assumptions to demonstrate the central idea of the methodology, we will address future studies toward application to real galaxy catalogs, as summarized below.

\begin{figure}
    \centering
    \includegraphics[scale=0.52]{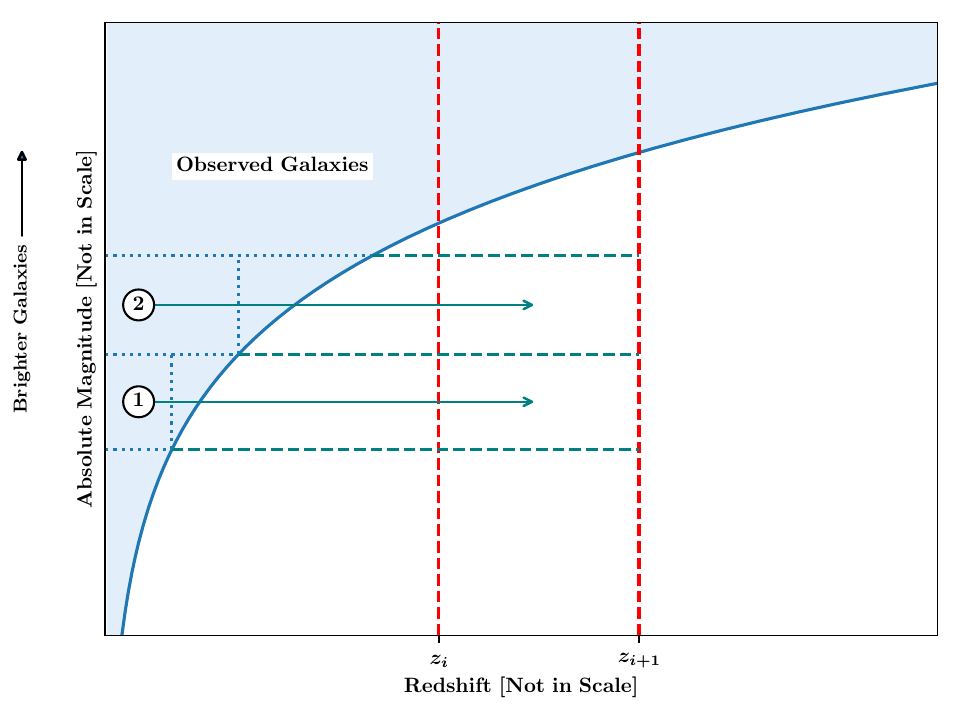}
    \caption{Schematic illustration of the proposed calibration scheme for $\brel$ in future applications to real data. The missing galaxy population in the target redshift bin $[z_i,z_{i+1})$ is divided into several absolute-magnitude bins, represented schematically by bins $1$ and $2$. At lower redshifts, galaxies in each corresponding magnitude bin and the brighter observed population are both observable, allowing $\brel$ to be measured directly from their clustering. The measured $\brel$ can then be extrapolated to the corresponding magnitude bin in the target redshift bin, as shown by the arrows.}
    \label{fig:M_z_schematic_future}
\end{figure}

\begin{itemize}

    \item [$-$] As discussed earlier, we compute $\brel$ using the underlying halo distribution while assuming the fiducial cosmology of the BigMDPL simulation. In practice, $\brel$ can instead be inferred from the observed clustering of analogous galaxy populations at lower redshifts. Specifically, as illustrated in Fig.~\ref{fig:M_z_schematic_future}, the missing population within the target redshift bin $(z_i, z_{i+1})$ can be divided into absolute magnitude bins. Each magnitude bin remains observable at some lower redshift, where its relative bias can be measured directly from the clustering of the corresponding bright galaxies. This measured value is then extrapolated to the target redshift bin. Such an extrapolation will require an assumed model, but this model is less restrictive and more directly motivated by the astrophysical scaling of galaxy bias with luminosity and redshift. This contrasts with approaches that explicitly model the galaxy or matter density field or their two-point correlation function. Since $\brel$ also depends on the assumed cosmology, this dependence must be accounted for as well. This can be achieved by first reconstructing the missing population for a fiducial cosmology; the resulting samples can then be reweighted as the cosmological parameters are varied during inference, with the weights calibrated in advance from the cosmology dependence of $\brel$.

    \item [$-$] Another simplification adopted in this study is the direct computation of $\nbmiss$ from the missing galaxy population itself. In practice, this quantity should be estimated in each redshift bin by integrating the Schechter luminosity function below the survey luminosity threshold, corresponding to the minimum observable luminosity, to obtain the number density of the missing galaxies.

    \item [$-$] A further avenue for improving the reconstruction for the more realistic Catalog C is the use of an adaptive HEALPix pixelization. Since the fraction of missing galaxies increases with redshift, the angular resolution can be chosen to vary with redshift, with finer pixels at low redshifts and hierarchically combined pixels at higher redshifts to maintain a sufficient number of observed galaxies in each pixel--redshift bin.
 
\end{itemize}

Addressing these assumptions will enable us to validate the framework for dark siren cosmology using mock galaxy catalogs and to compare its performance with that of the current framework, which neglects galaxy clustering. This comparison will be the immediate focus of our future work, providing a direct assessment of the impact of accounting for galaxy clustering in dark siren analyses.

\acknowledgments

T.G. acknowledges the use of the LDG cluster Sarathi at IUCAA for computations performed in this work and support from the JSPS Grant-in-Aid for Transformative Research Areas (A) No. 23H04893.
T.G. is also grateful to the Inter-University Centre for Astronomy and Astrophysics for its support during a research visit, where part of this work was carried out.

The MultiDark Database used in this paper and the web application providing online access to it were constructed as part of the activities of the German Astrophysical Virtual Observatory as a result of a collaboration between the Leibniz-Institute for Astrophysics Potsdam (AIP) and the Spanish MultiDark Consolider Project No. CSD2009-00064. The  Big MultiDark Planck simulation has been performed in the Supermuc supercomputer at LRZ using time granted by PRACE.

\bibliography{reference}

\end{document}